\documentclass[3p,times,twocolumn]{elsarticle}

\usepackage{ecrc}
\usepackage{amsmath}
 
\volume{00}

\firstpage{1}

\journalname{Proceedings of the Neutrino 2012 Conference, June 3-9,
  2012, Kyoto, Japan}

\runauth{C.D.~Ott et al.}

\jid{nuphbp}

\usepackage{amssymb}

\usepackage{color}

\def\apj{Astrophys. J.}

\def\apjs{Astrophys. J. Supp. Ser. }

\def\aap{Astron. Astrophys. }

\def\physrep{Phys. Rep. }

\def\prl{Phys. Rev. Lett.}
\def\prd{Phys. Rev. D.}

\begin{document}
\begin{frontmatter}



\dochead{}

\title{Core-Collapse Supernovae, Neutrinos, and Gravitational Waves}

\author[label1,label2]{C.~D.~Ott}
\author[label3]{E.~P.~O'Connor}
\author[label1]{S. Gossan}
\author[label1]{E.~Abdikamalov}
\author[label1]{U. C. T. Gamma}
\author[label4,label1]{S. Drasco}
\address[label1]{TAPIR, California Institute of Technology, Pasadena, California, USA}
\address[label2]{Kavli Institute for the Physics 
  and Mathematics of the Universe, Kashiwa, Chiba, Japan}
\address[label3]{Canadian Institute for Theoretical Astrophysics, Toronto, Ontario, Canada}
\address[label4]{Grinnell College, Grinnell, Iowa, USA}

\begin{abstract}
Core-collapse supernovae are among the most energetic cosmic
cataclysms. They are prodigious emitters of neutrinos and quite likely
strong galactic sources of gravitational waves.  Observation of both
neutrinos and gravitational waves from the next galactic or near
extragalactic core-collapse supernova will yield a wealth of
information on the explosion mechanism, but also on the structure and
angular momentum of the progenitor star, and on aspects of fundamental
physics such as the equation of state of nuclear matter at high
densities and low entropies. In this contribution to the proceedings
of the Neutrino 2012 conference, we summarize recent progress made in
the theoretical understanding and modeling of core-collapse
supernovae. In this, our emphasis is on multi-dimensional processes
involved in the explosion mechanism such as neutrino-driven convection
and the standing accretion shock instability. As an example of how
supernova neutrinos can be used to probe fundamental physics, we
discuss how the rise time of the electron antineutrino flux observed
in detectors can be used to probe the neutrino mass
hierarchy. Finally, we lay out aspects of the neutrino and
gravitational-wave signature of core-collapse supernovae and discuss
the power of combined analysis of neutrino and gravitational wave data
from the next galactic core-collapse supernova.
\end{abstract}

\end{frontmatter}


\section{Introduction}
\label{sec:introduction}

Core-collapse supernovae (CCSNe), primarily due to their
nucleosynthetic impact on the chemical evolution of the universe, are
of great importance in astrophysics. Their inner workings or, more
precisely, the physical processes that convert the necessary fraction
of gravitational energy of collapse into energy of the explosion,
involve a broad range of physics, in particular neutrino, nuclear, and
gravitational physics.  The cores of CCSNe are thus cosmic
laboratories for fundamental physics that can be probed by the
observation and analysis of neutrinos and gravitational-wave signals
from galactic or near-extragalactic CCSNe. This was first and
impressively demonstrated by the observation of neutrinos from SN
1987A \cite{hirata:87,bionta:87}, which confirmed the very basics of
CCSN theory \cite{bethe:90}: CCSNe are driven by the release of
gravitational energy in the collapse of a massive star's core and the
vast majority ($\sim$$99\%$) of this energy is liberated in
$\mathrm{MeV}$ neutrinos.

CCSN theory and modeling have come a long way since the first
pioneering numerical studies carried out in the late 1960s and early
1970s \cite{colgate:66,arnett:66,ivanova:69,wilson:71}.  Advances in
nuclear and neutrino physics, in combination with improvements in
numerical modeling methodology and rapid advances in computer
technology have since enabled detailed and self-consistent
spherically-symmetric (1D), axisymmetric (2D), and the first (3D)
models of CCSNe (see \cite{janka:12b,burrows:12b,kotake:12snreview}
for reviews).

It is now clear that the prompt hydrodynamic mechanism proposed early
on \cite{colgate:60} fails robustly. The shock created at core bounce
always stalls due to dissociation of infalling heavy nuclei and
neutrino losses from the region behind the shock. It must be revived
by the \emph{explosion mechanism}. The (delayed) neutrino mechanism
\cite{bethewilson:85} is now believed to drive most CCSN explosions
\cite{janka:12b}.  It relies on the deposition behind the shock
(resulting in net energy gain) of a small fraction ($\sim$$10\%$) of
the outgoing $\nu_e$ and $\bar{\nu}_e$ neutrino luminosity within a
few hundred milliseconds after core bounce. The neutrino mechanism has
been shown to work robustly in spherical symmetry for the lowest-mass
massive stars (e.g., \cite{kitaura:06}). Driving
explosions in more massive stars with the neutrino mechanism appears
to require an interplay of neutrino heating with multi-D fluid
dynamics due to neutrino-driven convection and the standing accretion
shock instability (SASI) (e.g.,
\cite{murphy:08,marek:09,mueller:12a,mueller:12b,nordhaus:10,hanke:12,murphy:12,dolence:12,burrows:12,ott:12b,bruenn:13}
and references therein). Convection/SASI leave characteristic imprints
on the neutrino and gravitational-wave signals, which, in turn, can be
used to observationally probe these instabilities
\cite{ott:09,marek:09b,murphy:09,lund:12,mueller:e12,mueller:12c}.

While garden-variety CCSN explosions (of spectral type II, Ib, Ic) may
very well be ultimately explained by the neutrino mechanism,
neutrino-driven explosions appear unfit to deliver hyper-energetic
explosions (with energies of $\gtrsim 10\,\mathrm{B}$; 1~[B]ethe $=
10^{51}\,\mathrm{erg}$). Such \emph{hypernovae} are predominantly of
the rare Ic-bl subclass (bl stands for relativistically Doppler
broadened spectral lines) of CCSNe that has also been associated with
long-duration gamma-ray bursts (e.g., \cite{modjaz:11}).  Such
hyperenergetic CCSN explosions may result from the magnetorotational
mechanism.  This mechanism relies on rapid progenitor star rotation
and efficient amplification of protoneutron star magnetic fields to
magnetar strength ($\sim$$10^{15}\,\mathrm{G}$) after bounce,
launching bipolar jet-like outflows
(e.g., \cite{bisno:70,burrows:07b}). The extreme
rotation required in this mechanism may leave characteristic imprints
on the neutrino and gravitational wave signal
\cite{ott:08,ott:12a}.

An interesting potential twist to the standard CCSN story developed
with the (re-)discovery of collective neutrino oscillation induced by
neutrino-neutrino coherent forward scattering (see, e.g.,
\cite{duan:10} for a review). Supernova
neutrinos, as reviewed by Balantekin in this volume, may oscillate in
vacuum, due to the MSW effect \cite{Mikheev:1986gs}, and, in regions
of high neutrino density, due to collective oscillations.  In the
inverted neutrino mass hierarchy, collective oscillations are
predicted to lead to a nearly complete swap of $\nu_e$/$\bar{\nu}_e$
spectra with the spectra of heavy lepton ($x = \mu, \tau$) neutrinos
$\nu_x$/$\bar{\nu}_x$. If this swap occurs below the region in which
neutrino heating dominates over neutrino cooling, this would
potentially lead to a boost of neutrino energy deposition, since the
emission spectra of $\nu_x/\bar{\nu}_x$ are significantly harder than
those of their $\nu_e/\bar{\nu}_e$ counterparts and the
charged-current absorption cross sections scale with the square of the
neutrino energy ($\epsilon_\nu^2$). That this could indeed have an
important impact on neutrino-driven explosions was first shown by
Suwa~et~al.~\cite{suwa:11a}. However, recent detailed oscillation
calculations on the basis of realistic supernova background matter
distributions and neutrino radiation fields found that collective
oscillations are likely to be suppressed in the early post-core-bounce
evolution and thus cannot enhance neutrino heating
\cite{chakraborty:11prl,sarikas:12,dasgupta:12a} (but see also
\cite{cherry:12}).

In this contribution to the proceedings of the Neutrino 2012
conferences, we outline, in Section~\ref{sec:multiD}, some recent
results from multi-D simulations of CCSNe. In Section~\ref{sec:nus},
we discuss how the early postbounce CCSN neutrino signal can be used
to constrain the neutrino mass hierarchy. Finally, in
Section~\ref{sec:next} we give examples of what may be learned from
combined observations of neutrinos and gravitational waves from the
next nearby CCSN.

\section{Results from Recent Neutrino-Driven 
Multi-Dimensional Core-Collapse Supernova Models}
\label{sec:multiD}

The first set of detailed 2D CCSN simulations carried out in the mid
1990s (e.g., \cite{herant:94,bhf:95,janka:96}) demonstrated that
the negative entropy gradient created by neutrino heating leads to
vigorous convection in the \emph{gain region} behind the shock (where
neutrino heating dominates over cooling). The added degree of freedom
in 2D liberates the infalling gas to move not just in the radial
direction (as in 1D), but also laterally. As a consequence, the dwell
time in the gain region of a given accreting mass element is
increased, leading to greater total neutrino heating and high-entropy
turbulent flow that pushes the shock to larger radii and aids the
explosion.

The standing accretion shock instability (SASI; e.g.,
\citep{blondin:03,scheck:08,foglizzo:06}) is another
hydrodynamic instability of relevance in the postbounce evolution of
CCSNe. In 2D simulations, it drives large-scale periodic low-order
($\ell = 1$ is the fastest growing mode) deformations of the shock
front, which, in the nonlinear phase of the instability, lead to
larger average shock radii, secondary shocks, and also increase the
dwell time of material in the gain region (e.g.,
\cite{scheck:08,murphy:08,mueller:12a,mueller:12b}). In 3D,
non-axisymmetric ($|m| > 0$) modes will develop. If rotation is present,
$m \pm 1$ spiral modes may appear (e.g., \cite{iwakami:09}), while in
the nonrotating case all $m$ of a given $\ell$ are degenerate and
should grow to comparable amplitudes, effectively spreading the
typical 2D $\ell$ amplitude over its $m$-modes in 3D
\cite{iwakami:08,burrows:12,dolence:12,ott:12b}.

There has been much recent discussion in the literature about the
respective roles of neutrino-driven convection and SASI and about
which of the two instabilities is the primary instability (i.e., the
one that grows first and is dominant)
\cite{mueller:12a,mueller:12b,hanke:12,burrows:12,
dolence:12,murphy:12,ott:12b}.
While this is not the place to review this discussion in detail, it is
worth pointing out a few key aspects:

\begin{figure}
\centering
\includegraphics[width=\columnwidth]{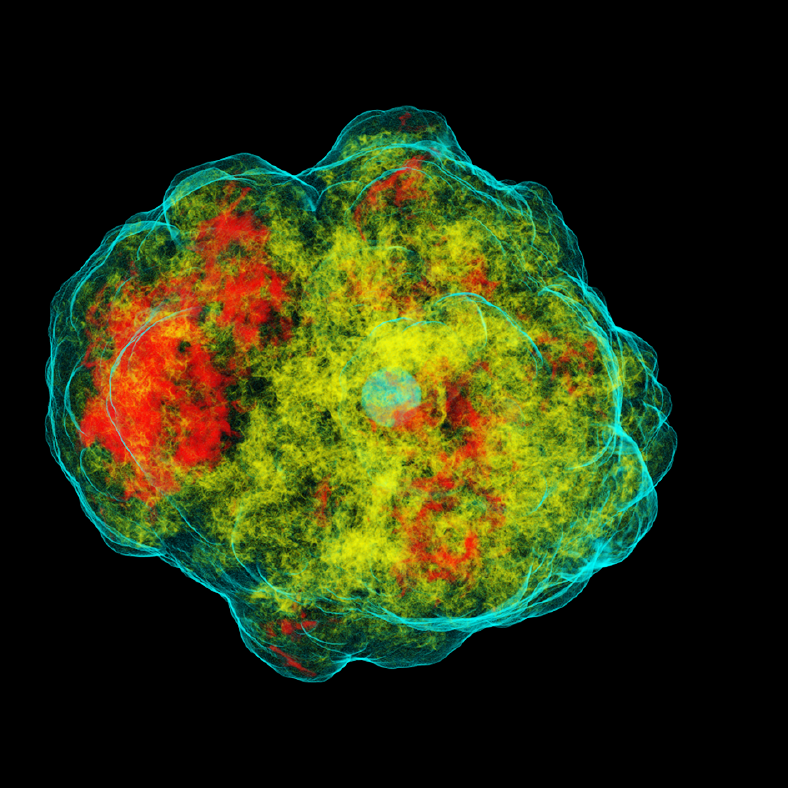}
\caption{3D volume rendering of the specific entropy $s$ at
  $\sim$$150\,\mathrm{ms}$ after core bounce in a 3D
  general-relativistic simulation carried out by
  Ott~et~al.~\cite{ott:12b}.  The scale of the frame is
  $700\,\mathrm{km}$ on a side. The colormap is chosen such that cyan
  correesponds to a moderate $s \sim
  4\,\mathrm{k_B}\,\mathrm{baryon}^{-1}$. Yellow and red indicate
  regions with $s \sim 16\,\mathrm{k_B}\,\mathrm{baryon}^{-1}$ and $s
  \sim 20\,\mathrm{k_B}\,\mathrm{baryon}^{-1}$, respectively.}
\label{fig:3dvol}
\end{figure}

(\emph{i}) Which instability grows first depends on (\emph{a}) the
early postbounce rate of advection through the region of convective
instability (and, hence, on the accretion rate set by progenitor star
structure) and (\emph{b}) on the size of the seed perturbations
present in the flow from which convection may grow
\cite{foglizzo:06,scheck:08} and thus on the level of asphericity in
the last stages of nuclear burning (e.g., \cite{arnett:11a}).  If the
initial size of a perturbation is too small to develop into a bouyant
plume before it is advected out of the region of convective
instability, then convection cannot grow and SASI will dominate (at
least initially) \cite{scheck:08,mueller:12b}.

(\emph{ii}) When the SASI approaches its nonlinear phase, it leads
to secondary shocks and seeds convection (e.g.,~\cite{scheck:08}), which,
in turn, may be limiting further SASI growth (e.g.,~\cite{guilet:10}).

(\emph{iii}) The behavior of the SASI appears fundamentally different
in 3D and 2D. In 2D, the dominant $\ell=1$ mode leads to strong
periodic north-south sloshing motions of the shock that persist even
when neutrino-driven convection becomes strong. In 3D, $\ell = 1$ is
still the fastest growing mode. However, its power is equipartitioned
into its $m = \{0,\pm1\}$ modes, the invidiual $(\ell,m)$ amplitudes
stay at much smaller values than in 2D and no large-scale single
sloshing mode develops
\cite{iwakami:08,nordhaus:10,mueller:e12,burrows:12,dolence:12,ott:12b}.

(\emph{iv}) In 3D, if neutrino-driven convection gets a chance to
grow, it appears to robustly become the dominant instability and
quench the SASI, in particular in simulations that have strong
neutrino heating and show explosions
\cite{burrows:12,dolence:12,ott:12b}. In the latter, non-oscillatory,
high-amplitude $\ell=1, m=\{0,\pm1\}$ deformations appear that
correspond to large bubbles of high-entropy that push out the shock
front in a particular direction \cite{burrows:12,dolence:12,ott:12b,couch:13a}.
An example of this is shown in Fig.~\ref{fig:3dvol}, which depicts
a volume rendering of the specific entropy at $\sim 150\,\mathrm{ms}$
after core bounce in a simulation of Ott~\emph{et al.}\cite{ott:12a}.

Nature is 3D. While highly detailed 2D multi-group
radiation-hydrodynamics simulations are now available
\cite{mueller:12a,mueller:12b}, a full understanding of CCSN
hydrodynamics must await equally detailed 3D simulations, which have
not been performed, but rapid progress towards them is being made (e.g.,
\cite{kuroda:12,takiwaki:12,ott:12b}).

\section{Supernova Neutrinos as Probes\\ of the Neutrino Mass Hierarchy}
\label{sec:nus}

Recently, Serpico~\emph{et al.} \cite{serpico:12} proposed that the
neutrino mass hierarchy could be determined from the early postbounce
neutrino signal of the next galactic CCSN for which complicated
collective oscillation effects may be ignored
\cite{chakraborty:11prl,sarikas:12}.  The original flavor (at
emission) of $\bar{\nu}_e$ arriving at a detector on Earth depends on
the neutrino mass hierarchy. For the normal hierarchy (NH), the MSW
effect \cite{Mikheev:1986gs} in the outer regions of the star leads to
an incomplete conversion of $\bar{\nu}_e$ to heavy-lepton
antineutrinos ($\bar{\nu}_x$) and vice versa.  For this hierarchy, in
the adiabatic approximation for neutrino oscillations, the
$\bar{\nu}_e$ signal at Earth is a linear combination of the original
$\bar{\nu}_e$ signal ($\propto \cos{(\theta_{12})}^2\sim 0.71$) and
the original $\bar{\nu}_x$ signal ($\propto
\sin{(\theta_{12})}^2 \sim 0.29$). However, for the inverted hierarchy
(IH) the $\bar{\nu}_e$ signal at Earth is completely composed of the
original $\bar{\nu}_x$ signal \cite{dighe:00}. So, for the
two hierarchies, we have:
\begin{eqnarray}
L^\mathrm{NH}_{\bar{\nu}_e,\oplus} &=& \cos{(\theta_{12})}^2 L_{\bar{\nu}_e,\mathrm{SN}} + 
\sin{(\theta_{12})}^2 L_{\bar{\nu}_x,\mathrm{SN}}\,\,,\label{eq:lumNH}\\
L^\mathrm{IH}_{\bar{\nu}_e,\oplus} &=& L_{\bar{\nu}_x,\mathrm{SN}}\,\,.\label{eq:lumIH}
\end{eqnarray}

\begin{figure}[t]
\centering
\includegraphics[width=\columnwidth]{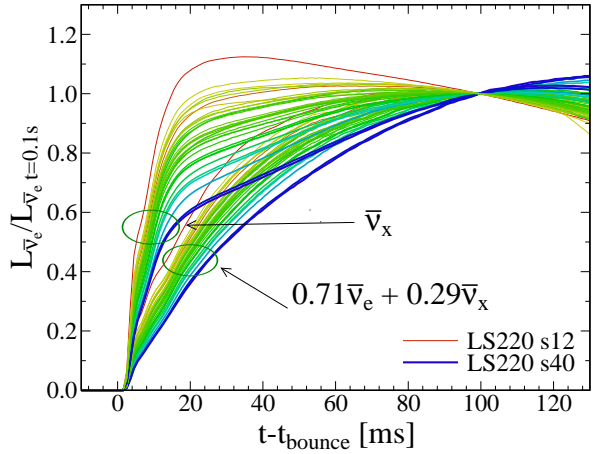}
\caption{Normalized (to the luminosity at 100\,ms after bounce)
  electron antineutrino ($\bar{\nu}_e$) luminosities at Earth for 32
  progenitor star models and assuming both the inverted (family of
  curves labeled via $\bar{\nu}_x$) and normal (family of curves
  labeled $0.71\bar{\nu}_e + 0.29\bar{nu}_x$) hierarchy. The colors
  distinguish progenitor models, colors ranging from red $\to$ yellow
  $\to$ green $\to$ blue denote increasing compactness,
  Eq.~(\ref{eq:bouncecompactness})
  \cite{oconnor:12a}. }\label{fig:normlums}
\end{figure}

\begin{figure}[t]
\centering \includegraphics[width=\columnwidth]{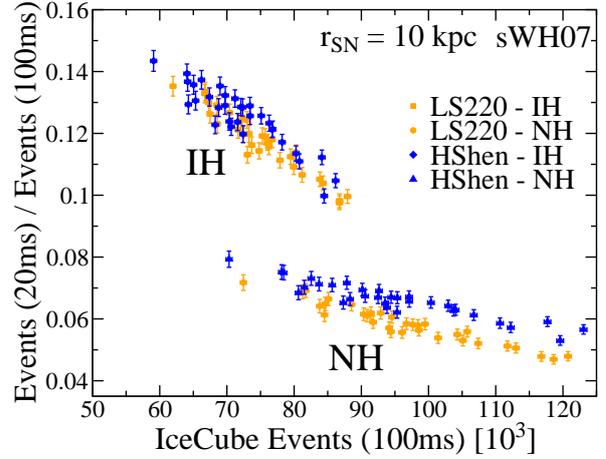}
\caption{Ratio of the number of expected events detected in the first
  20\,ms to the number of expected events detected in the first
  100\,ms versus cumulative events at 100\,ms. We show this ratio for
  each progenitor star model, EOS, and most important for our
  discussion, the assumed neutrino mass hierarchy. The fast rise time
  of the inverted hierarchy signal gives a large ratio for all models,
  independent of EOS and progenitor star details.  Note that neither
  axis begins at zero. Error bars are purely statistical and include
  the contribution from the IceCube background rate of
  $\sim$2200 events per 1.6384\,ms bin.}\label{fig:headtototal}
\end{figure}

The distinguishing characteristic explored in \cite{serpico:12}, and
further in this proceedings contribution, is the difference in the
rise times of the $\bar{\nu}_e$ signal and the $\bar{\nu}_x$
signal. The production of $\bar{\nu}_e$ is initially suppressed due to
the large electron chemical potential before bounce and in the very
early postbounce. It is not until the degeneracy has been partially
lifted in the outer regions of the protoneutron star that significant
$\bar{\nu}_e$ luminosity emerges from the supernova. Hence,
$\bar{\nu}_e$ have a slow rise time.  Heavy-lepton neutrino production
does not undergo this suppression and the $\bar{\nu}_x$ luminosity
rises much more quickly.  In the following, we explore whether a
high-event-rate detector like IceCube (e.g., \cite{icecube:11sn}) will
be able to distinguish between the two expected neutrino signals given
by eqs.~(\ref{eq:lumNH}) and (\ref{eq:lumIH}). In particular, we
extend the analysis of \cite{serpico:12} and explore whether there
exist degeneracies in the observed signal with progenitor star model
and equation of state (EOS).  We use results of O'Connor \& Ott
\cite{oconnor:12a}\footnote{Available at {\tt
    http://www.stellarcollapse.org/M1prog}} who simulate the
preexplosion phase in 32 solar-metallicity progenitor stars with
masses ranging from 12 to 120\,$M_\odot$ from
Ref.~\cite{woosley:07}. Two equations of state are used, the LS220 EOS
from \cite{lseos:91} and the HShen EOS from \cite{hshen:11}.  In
Fig.~\ref{fig:normlums}, we show the normalized (to its value at
100\,ms after bounce ) $\bar{\nu}_e$ luminosities expected at Earth
assuming both the normal (slowly rising curves) and inverted
hierarchies (fast rising curves).  We show only the luminosities for
the LS220 EOS.  The fast rise time of the inverted hierarchy is
clearly visible in the data. This figure is very similar to Fig.~4 of
\cite{serpico:12}, but shows a larger range of progenitors, which we
parameterize by their compactness parameter,
\begin{equation}
  \xi_{M} = {M / \,M_\odot \over R(M_\mathrm{bary} = M) /
    1000\,\mathrm{km}}\Big|_{t
    =t_{\mathrm{bounce}}}\,,\label{eq:bouncecompactness}
\end{equation}
where $R(M_\mathrm{bary}=M)$ is the radial coordinate that encloses a
baryonic mass of $M$ at the time of core bounce. Here, we use
$M=1.75\, M_\odot$, since this is a fiducial mass scale in the
postbounce preexplosion phase \cite{oconnor:12a}.  For clarity, the
color scale in Fig.~\ref{fig:normlums}, ranging from red $\to$ yellow
$\to$ green $\to$ blue, denotes increasing progenitor star compactness
\cite{oconnor:12a}.

We predict interaction rates in IceCube by using the following formula
\cite{ott:12a}, $R^\mathrm{IC}_{\bar{\nu}_e} \sim 2.6\times 10^4
\mathrm{s}^{-1} L_{\bar{\nu}_e,51}^\oplus/ r_{10\,\mathrm{kpc}}^2$,
where $L_{\bar{\nu}_e,51}^\oplus$ is the $\bar{\nu}_e$ luminosity at
Earth in units of $10^{51}$\,erg s$^{-1}$ and $r_{10\,\mathrm{kpc}}$
is the distances in units of 10\,kpc, the fiducial galactic
distance. We quantitatively discriminate between the two signals at
Earth by taking the fluence ratio of the number of events in the first
20\,ms to the number of events in the first 100\,ms.  While we have
not performed an analysis to find the optimal times to evaluate the
fluence ratio, 20\,ms/100\,ms is a reasonable choice, as one can see
in Fig.~\ref{fig:normlums}. It is apparent that by 20\,ms the
$\bar{\nu}_x$ luminosity has begun to level off. If we picked a time
smaller than 20\,ms, we would start to be dominated by statistical
error for the number of observed events. Times in the numerator longer
than 20\,ms lead to fluence ratios that are increasingly converging
towards one. The inverted hierarchy, which gives a faster rising
signal in $\bar{\nu}_e$ at Earth, will have a larger fluence ratio
than the normal hierarchy.  We show this ratio in
Fig.~\ref{fig:headtototal} for each progenitor model, EOS, and
hierarchy.  For clarity, we choose to plot the fluence ratio as a
function of total number of events\footnote{We note that unlike the
  ratio of events at 20\,ms to 100\,ms, the total number of events
  detected depends on the distance to the CCSN, which may not be
  determined (see \S\ref{sec:next}). However, given that the
  hierarchies are distinguishable for all models regardless of the
  total number of detected events (i.e. progenitor and EOS), this will
  not limit their distinguishability.} detected at 100\,ms. The
results of our extended model set confirm the results of obtained by
\cite{serpico:12}. The normal and inverted hierarchies should be
distinguishable from the rise time of the early neutrino signal from
the next galactic CCSN in an IceCube-like detector. In addition, over
the range of progenitors that we explore and the two quite distinct
EOS we employ, the signals predicted by the two hierarchies remain
clearly distinct, strengthening the ability to use supernovae neutrino
rise times as a probe of the neutrino mass hierarchy.

\section{The Next Galactic Core-Collapse Supernova: Gravitational Waves and Neutrinos Probing the Supernova Engine}
\label{sec:next}

The next galactic CCSN is guaranteed to come. While the predicted
rates for galactic CCSNe are in the range of a disheartening one per
$\sim$$30-100\,\mathrm{years}$ \cite{tammann:94,cappellaro:99}, we
prefer to look at this from a more positive perspective: \emph{The
  next galactic CCSN has already exploded.} Using the predicted range
of rates, one can, using a simple toy model of the galaxy assuming
uniform star formation in the disk, estimate that $\sim$$1000\,
(\mathrm{SN\,rate}) / (3.33\times
10^{-2}\,\mathrm{SNe}\,\mathrm{yr}^{-1})$ CCSNe have already exploded
and their photons, neutrinos, and gravitational waves (GWs) are on
their way to Earth.  The arrival of these messengers from the next
galactic CCSN may be imminent. Due to intervening dust and gas, a
significant fraction, perhaps $30-50\%$, of these CCSNe will not be
discovered in the optical, but will rather be seen as very luminous
infrared sources and/or via their X- and $\gamma$-ray emission.
Photons observed from CCSNe will initially carry information on the
outer layers of the progenitor star, and only at late times allow us
to probe central regions. Neutrinos and GW, on the other hand, will
provide plentiful information on the thermodynamics and dynamics of
the supernova engine.

As shown by Lund~\emph{et~al.} \cite{lund:12} (also seen in
\cite{brandt:11,ott:08,marek:09}), the SASI leads to characteristic
time variations in the neutrino flux at $\sim$$ 50-100\,\mathrm{Hz}$
that are observable by large Cherenkov detectors (see
\cite{scholberg:12} and other contributions in this volume for a
discussion of supernova neutrino detectors) for a galactic CCSN.  If
neutrino-driven convection dominates, the preexplosion time variations
of the neutrino flux may be expected to be characterized by smaller
amplitude and higher frequency.  Hence, the neutrino signal of the
next galactic CCSN may observationally constrain the prominence of
neutrino-driven convection and SASI.

GWs probe accelerated aspherical (of lowest-order quadrupole)
mass-energy motions. Detailed reviews of the CCSN GW signature can be
found in \cite{ott:09,kotake:12review}. In the case of a nonrotating
or slowly rotating CCSN, GW emission will occur via (\emph{i}) prompt
postbounce convection, which may grow on the negative entropy gradient
left behind by the stalling shock, (\emph{ii}) convection inside the
protoneutron star, (\emph{iii}) convection/SASI between the
protoneutron star and the stalled shock, (\emph{iv}) asymmetric
neutrino emission, and (\emph{v}) aspherical outflows.  The upcoming
second-generation of km-scale GW interferometers (Advanced LIGO
\cite{harry:10}, GEO-HF \cite{grote:10}, Advanced Virgo
\cite{virgostatus:11}, and KAGRA \cite{kagra:12}) and third-generation
instruments such as the Einstein Telescope \cite{punturo:10} will be most
sensitive to GWs in the frequency band $\sim$$30-1000\,\mathrm{Hz}$,
and will most likely detect GW emission from galactic and near
extragalactic CCSNe (within, perhaps,
$\sim$$100\,\mathrm{kpc}$). Emission processes (\emph{i}-\emph{iii})
lead to broadband emission in $\sim$$50-1000\,\mathrm{Hz}$, while
(\emph{iv}) and ($\emph{v}$) are characterized by low-frequency
($\lesssim 10\,\mathrm{Hz}$) GWs and are thus less relevant for
Earth-based detectors.

Studies investigating GW emission from convection/SASI (e.g.,
\cite{murphy:09,marek:09b,yakunin:10,mueller:e12,ott:12b,mueller:12c})
have shown that the strongest component of GW emission comes from the
deceleration of undershooting downflow plumes at the edge of the
protoneutron star. As the region of deceleration is changing in the
early postbounce phase, the time-frequency evolution of the GW signal
observed from the next galactic CCSN will provide information on the early
evolution of the structure, composition and thermodynamics of the
lower postshock region, set, primarily, by the nuclear EOS and the
postbounce accretion rate \cite{murphy:09,mueller:12c}. This information
encoded in GWs could be combined with information from neutrinos,
which also probe the structure of the outer protoneutron star
\cite{oconnor:12a}.

In the case of a rapidly spinning CCSN progenitor that may explode via
the magnetorotational mechanism, core collapse proceeds approximately
axisymmetrically \cite{ott:07prl,scheidegger:10b}.  At bounce, the
extreme acceleration experienced by the rotationally-deformed ($\ell =
2, m=0$) inner core leads to a strong burst of GWs, followed by
quasi-periodic ring-down oscillations which continue for
$\sim$$10-15\,\mathrm{ms}$. At later postbounce times, rotational
nonaxisymmetric ($m \ne 0$) dynamics may develop due to a shear
instability and lead to longer-term quasi-periodic GW emission at
typically twice the rotation frequency of the protoneutron star
\cite{ott:07prl,scheidegger:10b}. The characteristics of the GW signal
due to dynamically-relevant rotation are rather distinct from those of
the GW signal due to convection/SASI. If it is true that rapid
rotation leads to magnetorotational explosions, while slowly
rotating/nonrotating progenitor stars explode via the neutrino mechanism,
the GW signal alone may be sufficient to determine the explosion
mechanism via Bayesian model selection \cite{logue:12}.

The neutrino signal seen from a rapidly rotating CCSN will depend on
the inclination of the spin axis with respect to the observer (e.g.,
\cite{ott:08} and references therein): The rotational flattening of
the protoneutron star leads to higher neutrino fluxes with harder
spectra along the poles (i.e.~the rotation axis) and to lower fluxes
with softer spectra in equatorial regions.  The hardness of the
neutrino spectrum is also sensitive to the nuclear EOS (e.g.,
\cite{oconnor:12a,marek:09b}) and the inclination of the CCSN will not
be known, unless it goes along with a gamma-ray burst, indicating a
rotation axes aligned with the line of sight. Additional information
carried in the GW signal can be used to break the degeneracy between
rotation and EOS and constrain the unknown inclination angle.  For
example, if rotation is rapid, the protoneutron star ring-down
oscillations seen in the GW signal go along with time variations in
the $\bar{\nu}_e$ and $\nu_x$ fluxes at the same frequency that may be
detected out to $\mathcal{O}(1\,\mathrm{kpc})$ distances
\cite{ott:12a}.  If the CCSN spin axis is pointed at Earth, the $(\ell
= 2, m=0)$ GWs emitted by the ring-down oscillations will be
suppressed, but the time-variations in the neutrino fluxes will be
seen, as will be the subsequent GW emission by nonaxisymmetric
dynamics, which is of $(\ell=2,m=2)$ character and peaks along the
rotation axis \cite{scheidegger:10b}. Combining these different
emissions will allow observers to constrain the inclination of the
magnitude and orientation of the CCSN spin, in addition to helping to
break the degeneracy between rotation and nuclear EOS.

The potential of combined analysis of GW and neutrino signals from the
next galactic CCSN is tremendous. Much theoretical and modeling work
is still needed to flesh out the details and parameter dependence
(progenitor structure, rotation, EOS etc.) of both, GW and neutrino
signals. This will require further improvements in modeling technology
and efficiency to enable parameter studies in 3D. Progress is also
needed in the development of parameter estimation approaches that will
allow us to make statistical statements based on the
combination of information carried in GWs and neutrinos.

\section*{Acknowledgments}
We acknowledge helpful interactions at Neutrino 2012 with John Beacom,
Ryan Patterson, Kate Scholberg, and Mark Vagins. This research is
partially supported by NSF grant nos.\ AST-1212170, PHY-1151197,
PHY-1068881, and OCI-0905046, by the Alfred P.~Sloan Foundation, and by
the Sherman Fairchild Foundation.  Some of the results presented here
were obtained on supercomputers of the NSF XSEDE network under
computer time allocation TG-PHY100033.


\begin{thebibliography}{10}
\expandafter\ifx\csname url\endcsname\relax
  \def\url#1{\texttt{#1}}\fi
\expandafter\ifx\csname urlprefix\endcsname\relax\def\urlprefix{URL }\fi
\expandafter\ifx\csname href\endcsname\relax
  \def\href#1#2{#2} \def\path#1{#1}\fi

\bibitem{hirata:87}
K.~{Hirata}, T.~{Kajita}, M.~{Koshiba}, M.~{Nakahata}, Y.~{Oyama}, Phys. Rev.
  Lett. 58 (1987) 1490.

\bibitem{bionta:87}
R.~M. {Bionta}, G.~{Blewitt}, C.~B. {Bratton}, D.~{Casper}, A.~{Ciocio}, \prl
  58 (1987) 1494.

\bibitem{bethe:90}
H.~A. {Bethe}, Rev. Mod. Phys. 62 (1990) 801.

\bibitem{colgate:66}
S.~A. {Colgate}, R.~H. {White}, \apj 143 (1966) 626.

\bibitem{arnett:66}
W.~D. {Arnett}, Canadian Journal of Physics 44 (1966) 2553.

\bibitem{ivanova:69}
L.~N. {Ivanova}, V.~S. {Imshennik}, D.~K. {Nadezhin}, NInfo 13 (1969) 3.

\bibitem{wilson:71}
J.~R. {Wilson}, \apj 163 (1971) 209.

\bibitem{janka:12b}
H.-T. {Janka}, F.~{Hanke}, L.~{H\"udepohl}, A.~{Marek}, B.~{M\"uller},
  M.~{Obergaulinger}, Submitted to Prog. Th. Exp. Phys.; arXiv:1211.1378.

\bibitem{burrows:12b}
A.~{Burrows}, Rev. Mod. Phys. Collq. in press, arXiv:1210.4921.

\bibitem{kotake:12snreview}
K.~{Kotake}, K.~{Sumiyoshi}, S.~{Yamada}, T.~{Takiwaki}, T.~{Kuroda},
  Y.~{Suwa}, H.~{Nagakura}, Prog. Theo. Exp. Phys. 2012 (2012) 301.

\bibitem{colgate:60}
S.~A. {Colgate}, M.~H. {Johnson}, \prl 5 (1960) 235.

\bibitem{bethewilson:85}
H.~A. {Bethe}, J.~R. {Wilson}, \apj 295 (1985) 14.

\bibitem{kitaura:06}
F.~S. {Kitaura}, H.-T. {Janka}, W.~{Hillebrandt}, \aap 450 (2006) 345.

\bibitem{murphy:08}
J.~W. {Murphy}, A.~{Burrows}, \apj 688 (2008) 1159.

\bibitem{marek:09}
A.~{Marek}, H.-T. {Janka}, \apj 694 (2009) 664.

\bibitem{mueller:12a}
B.~{M{\"u}ller}, H.-T. {Janka}, A.~{Marek}, \apj 756 (2012) 84.

\bibitem{mueller:12b}
B.~{M{\"u}ller}, H.-T. {Janka}, A.~{Heger}, Submitted to the Astrophys. J.,
  ArXiv:1205.7078.

\bibitem{nordhaus:10}
J.~{Nordhaus}, A.~{Burrows}, A.~{Almgren}, J.~{Bell}, \apj 720 (2010) 694.

\bibitem{hanke:12}
F.~{Hanke}, A.~{Marek}, B.~{M{\"u}ller}, H.-T. {Janka}, \apj 755 (2012) 138.

\bibitem{murphy:12}
J.~W. {Murphy}, J.~C. {Dolence}, A.~{Burrows}, Submitted to the Astrophys. J.,
  ArXiv:1205.3491.

\bibitem{dolence:12}
J.~C. {Dolence}, A.~{Burrows}, J.~W. {Murphy}, J.~{Nordhaus}, Submitted to the
  Astrophys.~J., arXiv:1210.5241.

\bibitem{burrows:12}
A.~{Burrows}, J.~C. {Dolence}, J.~W. {Murphy}, \apj 759 (2012) 5.

\bibitem{ott:12b}
C.~D. {Ott}, E.~{Abdikamalov}, P.~{Moesta}, R.~{Haas}, S.~{Drasco},
  E.~{O'Connor}, C.~{Reisswig}, C.~{Meakin}, E.~{Schnetter}, Submitted to the
  Astrophys.~J., arXiv:1210.6674.

\bibitem{bruenn:13}
S.~W. {Bruenn}, A.~{Mezzacappa}, W.~R. {Hix}, E.~J. {Lentz}, O.~E. {Bronson
  Messer}, E.~J. {Lingerfelt}, J.~M. {Blondin}, E.~{Endeve}, P.~{Marronetti},
  K.~N. {Yakunin}, Submitted to Astrophys. J. Lett., arxiv:1212.1747.

\bibitem{ott:09}
C.~D. {Ott}, Class. Quantum Grav. 26 (2009) 063001.

\bibitem{marek:09b}
A.~{Marek}, H.-T. {Janka}, E.~{M{\"u}ller}, \aap 496 (2009) 475.

\bibitem{murphy:09}
J.~W. {Murphy}, C.~D. {Ott}, A.~{Burrows}, \apj 707 (2009) 1173.

\bibitem{lund:12}
T.~{Lund}, A.~{Wongwathanarat}, H.-T. {Janka}, E.~{M{\"u}ller}, G.~{Raffelt},
  \prd 86  (2012) 105031.

\bibitem{mueller:e12}
E.~{M{\"u}ller}, H.-T. {Janka}, A.~{Wongwathanarat}, \aap 537 (2012) A63.

\bibitem{mueller:12c}
B.~{M\"uller}, H.-T. {Janka}, A.~{Marek}, Submitted to the Astrophys.~J.,
  arxiv:1210.6984.

\bibitem{modjaz:11}
M.~{Modjaz}, Astron. Nachr. 332 (2011) 434.

\bibitem{bisno:70}
G.~S. {Bisnovatyi-Kogan}, Astron. Zh. 47 (1970) 813.

\bibitem{burrows:07b}
A.~{Burrows}, L.~{Dessart}, E.~{Livne}, C.~D. {Ott}, J.~{Murphy}, \apj 664
  (2007) 416.

\bibitem{ott:08}
C.~D. {Ott}, A.~{Burrows}, L.~{Dessart}, E.~{Livne}, \apj 685 (2008) 1069.

\bibitem{ott:12a}
C.~D. {Ott}, E.~{Abdikamalov}, E.~{O'Connor}, C.~{Reisswig}, R.~{Haas},
  P.~{Kalmus}, S.~{Drasco}, A.~{Burrows}, E.~{Schnetter}, \prd 86  (2012)
  024026.

\bibitem{duan:10}
H.~{Duan}, G.~M. {Fuller}, Y.-Z. {Qian}, Ann. Rev. Nuc. Part. Sc. 60 (2010)
  569.

\bibitem{Mikheev:1986gs}
S.~Mikheev, A.~Smirnov, Sov.J.Nucl.Phys. 42 (1985) 913.

\bibitem{suwa:11a}
Y.~Suwa, K.~Kotake, T.~Takiwaki, M.~{Liebend\"orfer}, K.~Sato, \apj 738 (2011)
  165.

\bibitem{chakraborty:11prl}
S.~{Chakraborty}, T.~{Fischer}, A.~{Mirizzi}, N.~{Saviano}, R.~{Tom{\`a}s},
  Phys. Rev. Lett. 107 (2011) 151101.

\bibitem{sarikas:12}
S.~{Sarikas}, G.~G. {Raffelt}, L.~{H{\"u}depohl}, H.-T. {Janka}, \prl 108
  (2012) 061101.

\bibitem{dasgupta:12a}
B.~{Dasgupta}, E.~P. {O'Connor}, C.~D. {Ott}, \prd 85 (2012) 065008.

\bibitem{cherry:12}
J.~F. {Cherry}, J.~{Carlson}, A.~{Friedland}, G.~M. {Fuller}, A.~{Vlasenko},
  \prl 108  (2012) 261104.

\bibitem{herant:94}
M.~{Herant}, W.~{Benz}, W.~R. {Hix}, C.~L. {Fryer}, S.~A. {Colgate}, \apj 435
  (1994) 339.

\bibitem{bhf:95}
A.~{Burrows}, J.~{Hayes}, B.~A. {Fryxell}, \apj 450 (1995) 830.

\bibitem{janka:96}
H.-T. {Janka}, E.~{M\"uller}, \aap 306 (1996) 167.

\bibitem{blondin:03}
J.~M. {Blondin}, A.~{Mezzacappa}, C.~{DeMarino}, \apj 584 (2003) 971.

\bibitem{scheck:08}
L.~{Scheck}, H.-T. {Janka}, T.~{Foglizzo}, K.~{Kifonidis}, \aap 477 (2008) 931.

\bibitem{foglizzo:06}
T.~{Foglizzo}, L.~{Scheck}, H.-T. {Janka}, \apj 652 (2006) 1436.

\bibitem{iwakami:09}
W.~{Iwakami}, K.~{Kotake}, N.~{Ohnishi}, S.~{Yamada}, K.~{Sawada}, \apj 700
  (2009) 232.

\bibitem{iwakami:08}
W.~{Iwakami}, K.~{Kotake}, N.~{Ohnishi}, S.~{Yamada}, K.~{Sawada}, \apj 678
  (2008) 1207.

\bibitem{arnett:11a}
W.~D. {Arnett}, C.~{Meakin}, \apj 733 (2011) 78.

\bibitem{guilet:10}
J.~{Guilet}, J.~{Sato}, T.~{Foglizzo}, \apj 713 (2010) 1350.

\bibitem{couch:13a}
S.~{Couch}, submitted to the \apj, arXiv:1212.0010.

\bibitem{kuroda:12}
T.~{Kuroda}, K.~{Kotake}, T.~{Takiwaki}, \apj 755 (2012) 11.

\bibitem{takiwaki:12}
T.~{Takiwaki}, K.~{Kotake}, Y.~{Suwa}, \apj 749 (2012) 98.

\bibitem{serpico:12}
P.~D. {Serpico}, S.~{Chakraborty}, T.~{Fischer}, L.~{H{\"u}depohl}, H.-T.
  {Janka}, A.~{Mirizzi}, \prd 85  (2012) 085031.

\bibitem{dighe:00}
A.~S. {Dighe}, A.~Y. {Smirnov}, \prd 62  (2000) 033007.

\bibitem{oconnor:12a}
E.~{O'Connor}, C.~D. {Ott}, Astrophys. J. in press, arXiv:1207.1100.

\bibitem{icecube:11sn}
R.~{Abbasi~et~al.\ [IceCube Collaboration]}, \aap 535 (2011) A109.

\bibitem{woosley:07}
S.~E. {Woosley}, A.~{Heger}, \physrep 442 (2007) 269.

\bibitem{lseos:91}
J.~M. Lattimer, F.~D. Swesty, {Nucl. Phys. A} 535 (1991) 331.

\bibitem{hshen:11}
H.~{Shen}, H.~{Toki}, K.~{Oyamatsu}, K.~{Sumiyoshi}, \apjs 197 (2011) 20.

\bibitem{tammann:94}
G.~A. {Tammann}, W.~{Loeffler}, A.~{Schroeder}, \apjs 92 (1994) 487.

\bibitem{cappellaro:99}
E.~{Cappellaro}, R.~{Evans}, M.~{Turatto}, \aap 351 (1999) 459.

\bibitem{brandt:11}
T.~D. {Brandt}, A.~{Burrows}, C.~D. {Ott}, E.~{Livne}, \apj 728 (2011) 8.

\bibitem{scholberg:12}
K.~{Scholberg}, Ann. Rev. Nuc. Part. Sc. 62 (2012) 81.

\bibitem{kotake:12review}
K.~{Kotake}, submitted to a special issue of Comptes Rendus Physique
  "Gravitational Waves (from detectors to astrophysics)", ArXiv:1110.5107.

\bibitem{harry:10}
G.~M. {Harry}, {LIGO Scientific Collaboration}, Class. Quantum Grav. 27  (2010)
  084006.

\bibitem{grote:10}
H.~{Grote}, {the LIGO Scientific Collaboration}, Class. Quantum Grav. 27 (2010)
  084003.

\bibitem{virgostatus:11}
T.~Accadia et~al.\ {(Virgo Collaboration)}, Class. Quantum Grav. 28  (2011)
  114002.

\bibitem{kagra:12}
K.~Somiya~{(for the KAGRA collaboration)}, Class. Quantum Grav. 29  (2012)
  124007.

\bibitem{punturo:10}
M.~{Punturo}, et~al., Class. Quantum Grav. 27 (2010) 194002.

\bibitem{yakunin:10}
K.~N. {Yakunin}, P.~{Marronetti}, A.~{Mezzacappa}, S.~W. {Bruenn}, C.-T. {Lee},
  M.~A. {Chertkow}, W.~R. {Hix}, J.~M. {Blondin}, E.~J. {Lentz}, O.~E. {Bronson
  Messer}, S.~{Yoshida}, Class. Quantum Grav. 27 (2010) 194005.

\bibitem{ott:07prl}
C.~D. {Ott}, H.~{Dimmelmeier}, A.~{Marek}, H.-T. {Janka}, I.~{Hawke},
  B.~{Zink}, E.~{Schnetter}, \prl 98 (2007) 261101.

\bibitem{scheidegger:10b}
S.~{Scheidegger}, R.~{K{\"a}ppeli}, S.~C. {Whitehouse}, T.~{Fischer},
  M.~{Liebend{\"o}rfer}, \aap 514 (2010) A51.

\bibitem{logue:12}
J.~{Logue}, C.~D. {Ott}, I.~S. {Heng}, P.~{Kalmus}, J.~{Scargill}, \prd 86
  (2012) 044023.

\end{thebibliography}
\end{document}